\documentstyle[12pt,psfig]{article}
\newcommand{\GeV}{~\mbox{GeV}}

\newcommand{\gsim}{ \mathop{}_{\textstyle \sim}^{\textstyle >} }
\newcommand{\lsim}{ \mathop{}_{\textstyle \sim}^{\textstyle <} }

\newcommand{\vev}[1]{ \left\langle {#1} \right\rangle }
\newcommand{\ep}{\epsilon}
\begin{document}
\baselineskip 0.6cm
\def\tr{\mathop{\rm tr}\nolimits}
\def\Tr{\mathop{\rm Tr}\nolimits}
\def\Re{\mathop{\rm Re}\nolimits}
\def\Im{\mathop{\rm Im}\nolimits}
\setcounter{footnote}{1}

\begin{titlepage}

\begin{flushright}
UT-924\\
hep-ph/0102187\\
\end{flushright}
 
\vskip 2cm
\begin{center}
 {\large \bf Reheating-temperature independence of cosmological
 baryon asymmetry in Affleck-Dine leptogenesis}

\vskip 1.2cm 

Masaaki Fujii$^1$, K. Hamaguchi$^1$, and T. Yanagida$^{1,2}$

 \vskip 0.4cm

 {\it $^1$ Department of Physics, University of Tokyo,
 Tokyo 113-0033, Japan}\\
 {\it $^2$ Research Center for the Early Universe, University of Tokyo,
 Tokyo 113-0033, Japan}
\vskip 0.2cm
\vskip 2cm
\abstract{In this paper we point out that the cosmological baryon
 asymmetry in our universe is generated almost independently of the
 reheating temperature $T_R$ in Affleck-Dine leptogenesis and it is
 determined mainly by the mass of the lightest neutrino, $m_{\nu_1}$, in
 a wide range of the reheating temperature $T_R\simeq 10^5$--$10^{12}$
 GeV.  The present baryon asymmetry predicts the $m_{\nu_1}$ in a narrow
 region, $m_{\nu_1}\simeq (0.3$--$1)\times 10^{-9}$ eV. \, Such a small
 mass of the lightest neutrino leads to a high predictability on the
 mass parameter $m_{\nu_e \nu_e}$ contributing to the neutrinoless
 double beta decay. We also propose an explicit model in which such an
 ultralight neutrino can be naturally obtained. }

\end{center}
\end{titlepage}

\setcounter{footnote}{0}

%
\section{Introduction}
The origin of baryon (matter-antimatter) asymmetry in the present
universe is one of the fundamental problems in particle physics as well
as in cosmology. Recently, leptogenesis~\cite{LGorg} becomes very
attractive among various baryogeneis scenarios, since there are now
convincing evidences of neutrino oscillations, especially the
atmospheric neutrino oscillation reported by the Superkamiokande
collaboration~\cite{SK-Atm}. The small but nonzero masses of neutrinos
suggested from neutrino-oscillation experiments strongly indicate the
existence of lepton-number violation, which is a crucial ingredient of
the leptogenesis.  It is extremely interesting in the leptogenesis
scenario that the baryon asymmetry of the present universe is closely
related to observable low-energy physics, namely, neutrino masses and
mixings.

Among various mechanisms~\cite{LG-Th,LG-ID,LG-AD,LG-AD-others} for
leptogenesis proposed so far, the leptogenesis of Affleck-Dine (AD)
mechanism~\cite{Affleck-Dine} is naturally expected to work once one
introduces supersymmetry (SUSY) in the standard model together with the
nonzero neutrino mass. Recently, a detailed analysis on this mechanism
(AD leptogenesis~\cite{LG-AD}) has been performed~\cite{AFHY}, which has
shown that the dynamics of the flat direction field is drastically
changed by thermal effects, as pointed out in
Refs.~\cite{DRT,ACE}. Actually, it has been shown~\cite{AFHY} that the
resultant baryon asymmetry is indeed suppressed for relatively high
reheating temperatures $T_R$, and an ultralight neutrino is required in
order to obtain the sufficient baryon asymmetry $n_B/s$ in the present
universe.  (Here $n_B$ and $s$ are the baryon number and entropy density
in the present universe, respectively.)

In this paper, we perform a reanalysis on the AD leptogenesis, and
emphasize that the present baryon asymmetry $n_B/s$ is determined mainly
by the mass of the lightest neutrino $m_{\nu_1}$ for $T_R\simeq
10^5$--$10^{12}$ GeV, and its dependence on the reheating temperature
$T_R$ is rather mild. Here, we include an additional thermal effect
observed recently in Ref.~\cite{SomeIssues}, which makes the dependence
of the resultant baryon asymmetry on the reheating temperature even
milder. Notice that in many baryogenesis scenarios the obtained baryon
asymmetry depends crucially on the reheating temperature $T_R$ of the
inflation. Thus, it is very attractive that the baryon asymmetry in our
universe is predicted almost independently of the reheating temperature
$T_R$ in the AD leptogenesis.

Furthermore, this reheating-temperature independence of the AD
leptogenesis means that the mass of the lightest neutrino $m_{\nu_1}$
can be determined from the present baryon asymmetry $n_B/s$. Actually,
we show that the observed baryon asymmetry $n_B/s\simeq (0.4$--$1)\times
10^{-10}$~\cite{KT} predicts $m_{\nu_1}\simeq (0.1$--$3)\times 10^{-9}$
eV. Thus, neutrinos can not be degenerate in mass, and the masses of the
two heavier neutrinos are also determined from the neutrino-oscillation
experiments for the atmospheric and the solar neutrinos, that is,
$m_{\nu_3} \simeq \sqrt{\delta m_{\rm atm}^2} \simeq (3$--$8) \times
10^{-2}$ eV~\cite{SK-Atm} and $m_{\nu_2} \simeq \sqrt{\delta m_{\rm
sol}^2} \sim 10^{-3}$--$10^{-2}$ eV~\cite{SK-Sol-recent}.

Although it is hard to confirm the mass of such an ultralight neutrino
$\nu_1$, it can be tested indirectly. A crucial observation here is that
such a small mass of the lightest neutrino together with the masses and
mixings of the neutrinos obtained from neutrino-oscillation experiments
suggests a high predictability on the rate of neutrinoless double beta
($0\nu\beta\beta$) decay. We show that the $\nu_e$--$\nu_e$ component of
the neutrino mass matrix, $m_{\nu_e \nu_e}$, contributing to the
$0\nu\beta\beta$ decay, is determined with high accuracy, depending on
the solution to the solar neutrino deficits and the $e$--$3$ component
of the neutrino mixing matrix, $U_{e3}$. For the case of large angle MSW
solution, which is favored from the recent Superkamiokande
experiments~\cite{SK-Sol-recent}, a sizable $m_{\nu_e \nu_e}$, say,
$|m_{\nu_e \nu_e}|\sim 10^{-2}$--$10^{-3}$ eV is predicted.  It is very
encouraging that such a large $|m_{\nu_e \nu_e}|$ is accessible at
future experiments for the $0\nu\beta\beta$ decay such as
GENIUS~\cite{GENIUS}. On the other hand, we find that the obtained
$|m_{\nu_e \nu_e}|$ depends highly on $U_{e3}$ for the case of small
angle MSW and LOW solution. We also stress that this predictability on
the $0\nu\beta\beta$ decay is a generic consequence of the mass
hierarchy $m_{\nu_1} \ll m_{\nu_{2,3}}$.

Finally, we propose an explicit model based on a Froggatt-Nielsen (FN)
mechanism~\cite{FroNie}, in which such an ultralight neutrino is
naturally predicted. Here, we impose a discrete $Z_6$ group as the FN
symmetry.

%
\section{Affleck-Dine leptogenesis}
\label{sec-AD}
The Affleck-Dine (AD) flat direction scalar field for leptogenesis
is~\cite{LG-AD}
\begin{eqnarray}
 \label{ADfield}
 H_u = L_i = \frac{1}{\sqrt{2}}\phi_i
  \,,
\end{eqnarray}
and along this direction we have effective dimension-five operators in
superpotential;
\begin{eqnarray}
 \label{Pot-super}
  W = \frac{1}{2M_i}\left(L_i H_u\right)^2
  = \frac{m_{\nu_i}}{2 \vev{H_u}^2} \left(L_i H_u\right)^2
  \,,
\end{eqnarray}
where $\vev{H_u} = \sin\beta \times 174$ GeV and $\tan\beta \equiv
\vev{H_u}/\vev{H_d}$. Here $H_u$ ($H_d$) and $L_i$ are Higgs field which
couple to up (down) type quarks and SU$(2)_L$-doublet lepton fields,
respectively. (They denote chiral superfields or their scalar
components.) Hereafter, we take $\sin\beta\simeq 1$.  Note that the
scale $M_i$ are related to the neutrino masses $m_{\nu_i}$ through the
seesaw formula $m_{\nu_i} = \vev{H_u}^2 / M_i$~\cite{seesaw}. Here we
have taken a basis where the mass matrix for neutrinos $\nu_i$ is
diagonal, and for simplicity we will suppress the family index $i$ ($=
1, 2, 3$) unless we denote it explicitly.

Before discussing the detailed dynamics of the $\phi$ field, we first
roughly describe the evolution of $\phi$ and note the relevant epoch for
the AD leptogenesis. After the end of inflation, the inflaton $\chi$
starts a coherent oscillation. (At this stage the energy density of the
universe is dominated by the oscillating inflaton $\chi$ and the Hubble
parameter $H$ of the expanding universe decreases with cosmic time $t$
as $H = (2/3)t^{-1}$~\cite{KT}.) After that, when the Hubble parameter
$H$ becomes comparable to the decay rate of the inflaton,
$\Gamma_{\chi}\sim T_R^2/M_*$ ($M_* = 2.4\times 10^{18}$ GeV is the
reduced Planck mass), the energy density of the radiation starts to
dominate the universe. The evolution of the $\phi$ field is as
follows. During the inflation, the $\phi$ field takes a large value
determined from the effective potential discussed below.  After the end
of inflation, the value of $\phi$ gradually decreases as the Hubble
parameter $H$ decreases and then, at some time, the $\phi$ starts its
coherent oscillation.  As we will see, the net lepton number is fixed
when the flat direction field $\phi$ starts the coherent oscillation.

Let us discuss the dynamics of the $\phi$ field. The method of the
following analysis in this section is based on Ref.~\cite{AFHY}.  First,
we show the total effective potential for the flat direction field
$\phi$ relevant to the leptogenesis. In addition to the usual $F$-term
potential and SUSY-breaking terms, there are additional SUSY-breaking
terms caused by the nonzero energy density of inflaton, which depend on
the Hubble parameter $H$, and also there are thermal potential terms
depending on the temperature $T$. It turns out that the total potential
is given by the following form;\footnote{The thermal potential
proportional to $T^4\log\left( |\phi|^2 \right)$ in
Eq.~(\ref{Pot-total}), which has been recently found out in
Ref.~\cite{SomeIssues}, was not considered in Ref.~\cite{AFHY}.}
\begin{eqnarray}
 \label{Pot-total}
  V_{\rm total} &=&
  \left( m_{\phi}^2 - H^2 + \sum_{f_k|\phi| < T} c_k f_k^2 T^2 \right)
  |\phi|^2
  \nonumber\\ &&
  + \frac{m_{3/2}}{8 M} \left(a_m \phi^4 + H.c.\right)
  + \frac{H}{8 M} \left(a_H \phi^4 + H.c.\right)
  \nonumber\\ &&
  + a_g \alpha_s^2 \,T^4\log\left(\frac{|\phi|^2}{T^2}\right)
  \nonumber\\ &&
  + \frac{1}{4 M^2}|\phi|^6
  \,.
\end{eqnarray}
We explain each terms by turns.  First of all, the $F$-term potential
directly comes from the superpotential Eq.~(\ref{Pot-super});
\begin{eqnarray}
 \label{Pot-F}
  V_F = \frac{1}{4 M^2}|\phi|^6
  \,.
\end{eqnarray}
Next, the SUSY-breaking potential for $\phi$ is given by
\begin{eqnarray}
 \delta V_0
  = m_{\phi}^2 |\phi|^2 
  + \frac{m_{3/2}}{8 M}\left(a_m \phi^4 + H.c.\right)
  \,,
\end{eqnarray}
where $m_{\phi}$ and $m_{3/2}a_m$ are SUSY-breaking mass parameters.
They are expected to be $m_{\phi}\sim m_{3/2}\sim 1$ TeV and $|a_m|\sim
1$.\footnote{In this paper we assume the gravity-meditation model of
SUSY-breaking.}

During the inflation and during the oscillation period of inflaton
$\chi$ after the end of inflation, the energy density of the universe is
dominated by the inflaton $\chi$. Thus, there appears additional
SUSY-breaking effects caused by the nonzero energy density of
$\chi$~\cite{DRT};
\begin{eqnarray}
 \label{Pot-inf}
  \delta V_{\rm inf} 
  = -c_H H^2 |\phi|^2 + \frac{H}{8 M}\left(a_H \phi^4 + H.c.\right)
  \,,
\end{eqnarray}
where $c_H$ and $a_H$ are real and complex constants,
respectively. Hereafter, we assume $c_H\simeq |a_H|\simeq 1$, since we
find that the conclusions in the present paper do not depend much on
these parameters unless $c_H \ll -1$.\footnote{See, for example,
Ref.~\cite{MM}.}

The rests in Eq.~(\ref{Pot-total}) correspond to the thermal effects
which we discuss now.  Although the energy density is dominated by the
inflaton $\chi$ during the inflaton-oscillation period, there is a
dilute plasma consisting of the decay products of the inflaton $\chi$
even in this period. The temperature of this dilute plasma is given
by~\cite{KT}
\begin{eqnarray}
 T = \left( T_R^2 M_* H \right)^{1/4} 
  \,.
\end{eqnarray}
This dilute plasma has crucial effects on the dynamics of the $\phi$
field. First, the fields $\psi_k$ which couple to $\phi$ are produced by
the inflaton decay and/or by thermal scatterings if their effective
masses are less than the temperature;
\begin{eqnarray}
 f_k |\phi| < T 
  \,,
\end{eqnarray}
and hence the flat direction field $\phi$ receives a thermal mass of
order $\sim f_k T$. Here, $f_k$ denote Yukawa or gauge coupling
constants of $\psi_k$ to $\phi$. The induced thermal mass term is given
by~\cite{ACE};
\begin{eqnarray}
 \delta V^{\rm th}_1 = \sum_{f_k|\phi| < T} c_k f_k^2 T^2 |\phi|^2 
  \,,
\end{eqnarray}
where $c_k$ are real positive constants of order unity. (Details for
$c_k$ and couplings $f_k$ relevant to the flat direction $\phi$ can be
seen in Ref.~\cite{AFHY}.)

\begin{figure}[t]
 \label{Fig-running}
 \centerline{\psfig{figure=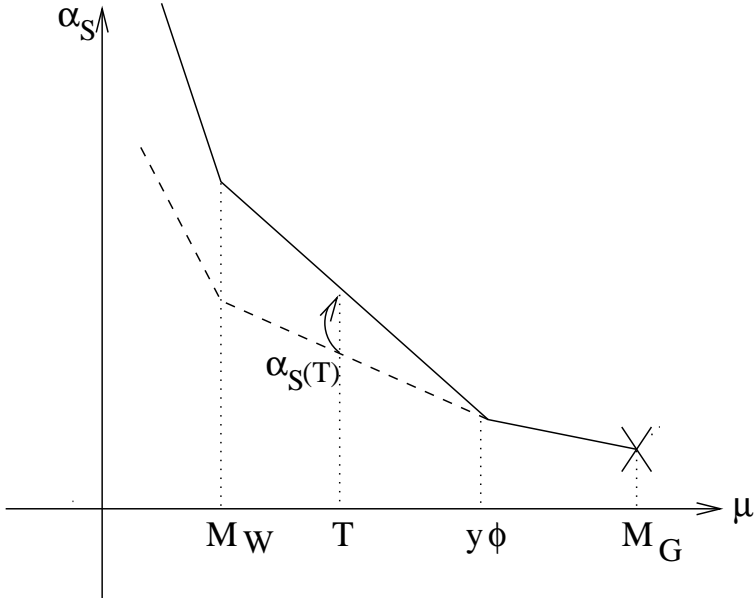,height=8cm}}
 \caption{A schematic behavior of the SU$(3)_C$ gauge coupling
 $\alpha_s$. The dashed line represents the running coupling when the
 $\phi$ field does not have a vacuum-expectation value.}
\end{figure}

Moreover, it has been recently pointed out~\cite{SomeIssues} that there
is another thermal effect. In the $\phi/\sqrt{2} = L = H_u$ flat
direction in Eq.~(\ref{ADfield}), the SU$(3)_C$ gauge group is not
broken, and gluons, gluinos and down-type (s)quarks remain
massless. These fields generate a free energy which is proportional to
$g_s(T)^2 T^4$ at two loop order,\footnote{In this flat direction, the
down type (s)quarks also generate a free energy of order
${\cal{O}}(y_{b}^{2})$, where $y_b$ is the Yukawa coupling of the bottom
quark. This gives an analogous free energy, which does not, however,
give a dominant effect as long as $y_{b}\lsim 1 $.} where $g_s$ is the
coupling constant of the SU$(3)_C$ gauge field. On the other hand, the
evolution of the running coupling $g_s$ is given by;
\begin{eqnarray}
 \frac{d}{d\,\log \mu}
  g_s(\mu)
  =
  \frac{g_s^3}{16 \pi^2}
  \left(
   - 3 C_2 
   + \sum_{f_i |\phi| < \mu}
   {\rm T}(R_i)
   \right)
   \,,
\end{eqnarray}
where $C_2 = 3$ and $T(R_i) = 1/2$ for fundamental
representations. Notice that the evolution changes when the scale $\mu$
passes through an effective mass of a field, $f_i|\phi|$, as shown in
Fig.~\ref{Fig-running}. In our case $f_i$ denote Yukawa couplings of
up-type quarks.
Thus, the coupling constant of SU$(3)_C$ depends on $|\phi|$ as follows;
\begin{eqnarray}
 \label{su3running}
  \left.
  g_s(T)
  \right|_{y_u |\phi| > T}
  =
  \left.
  g_s(T)
  \right|_{\phi = 0}
   +
   \frac{g_s(M_G)^3}{32\pi^2}
   \sum_{y_u |\phi| > T}
   {\rm T}(R_u)
   \log
   \left(
    \frac{y_u^2 |\phi|^2}{T^2}
    \right) 
    \,,
\end{eqnarray}
where $y_u$ are Yukawa coupling constants of up-type quarks and $M_G$ is
the ultraviolet scale where $g_s$ is fixed. Then, there is an additional
potential through the modification of the gauge coupling constant in
Eq.~(\ref{su3running});
\begin{eqnarray}
 \delta V^{\rm th}_2
  = a_g \alpha_s^2\,T^4
  \left(
   \sum_{y_u |\phi| > T}
   \frac{2}{3}
   {\rm T}(R_u)
   \right)
   \log\left(\frac{|\phi|^2}{T^2}\right) 
   \,,
\end{eqnarray}
where $a_g$ is a constant which is a bit larger than
unity~\cite{SomeIssues} and $\alpha_s \equiv g_s^2 / (4\pi)$. Hereafter,
we take the factor $\sum (2/3) T(R)$ to be unity since it does not
change the result much.\footnote{At least the top Yukawa coupling $f_t$
always satisfies $f_t|\phi| > T$ before the oscillation of $\phi$. Thus,
the resultant baryon asymmetry changes by only a factor of $\sqrt{3}$ at
most. See Eq.~(\ref{HO1}).}

Now we obtain the total effective potential Eq.~(\ref{Pot-total}), i.e.,
$V_{\rm total} = V_F + \delta V_0 + \delta V_{\rm inf} + \delta V^{\rm
th}_1 + \delta V^{\rm th}_2$. The evolution of $\phi$ is described by
the equation of motion with this $V_{\rm total}$ as
\begin{eqnarray}
 \label{EOM}
 \ddot{\phi} + 3 H \dot{\phi}
  + \frac{\partial V_{\rm total}}{\partial \phi^*} = 0 
  \,,
\end{eqnarray}
where the dot denotes a derivative with time.

During the inflation, there is no plasma and hence no thermal
effects. In this stage the Hubble parameter $H$ takes a constant value
which is much larger than the soft SUSY-breaking mass $m_{\phi}$. Thus,
the potential is dominated by the Hubble-induced terms $\delta V_{\rm
inf}$ in Eq.~(\ref{Pot-inf}) and $|\phi|^6$ term in Eq.~(\ref{Pot-F}), and
hence the flat direction $\phi$ rolls down to the minimum of the
potential,
\begin{eqnarray}
 \label{minima}
  |\phi| &\simeq& \sqrt{M H}
  \nonumber\\
 \arg(\phi) &\simeq& \frac{-\arg (a_H) + (2n+1) \pi}{4},\quad n = 0-3 
  \,.
\end{eqnarray}
Note that we have assumed the Hubble-induced mass squared is negative
($c_H \simeq +1 > 0$).

After the inflation ends, the inflaton $\chi$ starts to oscillate and
its decay produces a dilute plasma. However, the potential is still
dominated by Hubble-induced terms and $|\phi|^6$ term at the first stage
of the oscillation. Thus, the flat direction field $\phi$ is trapped for
a while in the above minimum given in Eq.~(\ref{minima})~\cite{DRT}.

Then, as the Hubble parameter decreases, the negative Hubble-induced
mass term is eventually exceeded by another term in the potential;
\begin{eqnarray}
 \label{preHO}
  H^2 \lsim m_{\phi}^2 
  + \sum_{f_k|\phi| < T} c_k f_k^2 T^2
  + a_g \alpha_s^2(T) \frac{T^4}{|\phi|^2} 
  \,.
\end{eqnarray}
As we shall see below, it is this time when the oscillation of $\phi$
starts. Let us denote the Hubble parameter at this time by $H_{\rm
osc}$. Using the relations $|\phi| \simeq \sqrt{M H}$ and $T = \left(
T_R^2 M_* H \right)^{1/4}$, it can be calculated from Eq.~(\ref{preHO})
as;
\begin{eqnarray}
 \label{HO1}
  H_{\rm osc} \simeq 
  \max
  \left[
   m_{\phi}
   \, ,\,
   H_i
   \, ,\,
   \alpha_s T_R \left(\frac{a_g M_*}{M}\right)^{1/2}
   \right] 
   \,,
\end{eqnarray}
where $H_i$ comes from the effect of the coupling $f_i$ and is given
by~\cite{AFHY}
\begin{eqnarray}
 \label{HO2}
  H_i \simeq \min
  \left\{
   \frac{1}{f_i^4}\frac{M_* T_R^2}{M^2}
   \, ,\,
   \left(c_i^2 f_i^4 M_* T_R^2 \right)^{1/3}
   \right\} 
    \,.
\end{eqnarray}

The evolution of the $\phi$ after $H \simeq H_{\rm osc}$ depends on
which term in Eq.~(\ref{Pot-total}) dominates the effective
potential. There are basically three cases; the potential is dominated
by (i) $m_{\phi}^2 |\phi|^2$ term, (ii) $T^2 |\phi|^2$ term, or (iii)
$T^4 \log(|\phi|^2)$ term.  First, if the potential is dominated by the
$m_{\phi}^2|\phi|^2$ term, the equation of motion Eq.~(\ref{EOM}) is
given by
\begin{eqnarray}
 \label{EOMmass}
  \ddot{\phi} + 3 H \dot{\phi}
  + m_{\phi}^2 \phi = 0 
  \,.
\end{eqnarray}
It is clear that the field $\phi$ oscillates around the origin ($\phi =
0$) and the amplitude of the oscillation dumps as $|\phi|\propto
H\propto t^{-1}$. Second, when the potential is dominated by the thermal
mass term $c_k f_k^2 T^2 |\phi|^2$, the $\phi$ field oscillates around
$\phi = 0$ and the amplitude dumps as $|\phi|\propto H^{7/8}\propto
t^{-7/8}$ \cite{AFHY}. The third case is given when the $a_g \alpha_s^2
T^4 \log (|\phi|^2/T^2)$ term dominates the potential. If we neglect the
time dependence of $T^4$, the damping rate of the oscillation amplitude
due to such a flat potential, $V\sim \log (|\phi|^2)$, can be estimated
by using the virial theorem, and it is given by $|\phi|\propto
H^2\propto t^{-2}$~\cite{GoMoMu}. In the actual case, however, the
potential itself gradually decreases with time as $T^4\propto
t^{-1}$. We have numerically checked that the amplitude dumps as
$|\phi|\propto H^{\alpha} \propto t^{-\alpha}$ with $\alpha\simeq
1.5$. Notice that, in all above cases, the dumping rate is faster than
the rate before the beginning of the $\phi$'s oscillation
($|\phi|\propto H^{1/2} \propto t^{-1/2}$).

Finally, we derive the resultant lepton asymmetry generated by the AD
leptogenesis mechanism. Since the $\phi$ field carries lepton charge,
its number density is related to the lepton number density $n_L$ as
\begin{eqnarray}
 \label{nL}
  n_L  = \frac{1}{2}i\left( \dot{\phi}^* \phi - \phi^* \dot{\phi}\right) 
  \,.
\end{eqnarray}
From Eqs.~(\ref{Pot-total}), (\ref{EOM}) and (\ref{nL}), the evolution of
$n_L$ is given by
\begin{eqnarray}
 \label{EOMnL}
 \dot{n}_L + 3 H n_L
  = \frac{m_{3/2}}{2 M}{\rm Im}\left(a_m \phi^4\right)
  + \frac{H}{2 M}{\rm Im}\left(a_H \phi^4\right) 
  \,.
\end{eqnarray}
A nontrivial motion of $\phi$ along the phase direction can generate a
net lepton asymmetry\cite{Affleck-Dine}. Although the $\phi$ field
almost traces one of the valleys in Eq.~(\ref{minima}), the phase of
$\phi$ is slightly kicked by the relative phase between $a_m$ and $a_H$
in the total potential in Eq.~(\ref{Pot-total}) during its rolling
towards the origin.  Thus, the first term in Eq.~(\ref{EOMnL}) plays a
role of the source of the lepton asymmetry.\footnote{The contribution to
the lepton asymmetry from the second term in Eq.~(\ref{EOMnL}) is always
less than or comparable to that from the first term, since ${\rm Im}(a_H
\phi^4)$ is suppressed. See Eq.~(\ref{minima}).}By integrating
Eq.~(\ref{EOMnL}), we obtain the resultant lepton asymmetry at the time
$t$,
\begin{eqnarray}
 \label{lepton-integral}
  \left[ R^3 n_L \right] (t)
  &\simeq& 
  \frac{m_{3/2}}{2 M} \int^t dt' R^3 
  {\rm Im}\left(a_m \phi^4 \right) 
  \,,
\end{eqnarray}
where $R$ denotes the scale factor of the expanding universe, which
scales as $R^3\propto H^{-2} \propto t^2$ in the universe dominated by
the oscillation energy of the inflaton $\chi$. We can see that the total
lepton number increases with time as $R^3 n_L \propto t$ until the
oscillation of $\phi$ starts ($H > H_{\rm osc}$), since $\phi^4 \propto
H^2$ and hence $R^3 \phi^4 \sim const$ in this stage. On the other hand,
after the $\phi$ starts its oscillation, the production of lepton number
is strongly suppressed. This is because ${\rm Im}\left(a_m \phi^4
\right)$ changes its sign rapidly due to the oscillation of $\phi$, and
also because the amplitude of $\phi$'s oscillation damped as $R^3
\phi^4\sim t^{-n}$ with $n > 1$. [See discussion below
Eq.~(\ref{EOMmass}).] Therefore, as mentioned in the beginning of this
section, the net lepton asymmetry is fixed when the oscillation of
$\phi$ starts. The generated lepton number at this epoch is given
approximately by
\begin{eqnarray}
 \label{nL-final}
  n_L =
  \left.
   \frac{m_{3/2}}{2 M}
   {\rm Im}\left(a_m \phi^4\right)
   t\,
   \right|_{H = H_{\rm osc}}
   =
   \frac{1}{3} m_{3/2} M H_{\rm osc} \delta_{\rm eff} 
   \,,
\end{eqnarray}
where $\delta_{\rm eff}\simeq \sin(4\arg\phi + \arg a_m)$ represents an
effective $CP$ violating phase. From Eq.~(\ref{nL-final}), the
lepton-to-entropy ratio is estimated as\footnote{In deriving
Eq.~(\ref{nLs-final}) we have assumed that the lepton number is fixed
before the reheating process of the inflation completes, namely, $H_{\rm
osc} \gsim T_R^2 / M_*$, which is satisfied as long as $T_R \lsim
10^{17}$ GeV for $m_{\nu}\lsim 10^{-6}$ eV.}
\begin{eqnarray}
 \label{nLs-final}
  \frac{n_L}{s} = \frac{M T_R}{12 M_*^2}
  \left(
   \frac{m_{3/2}}{H_{\rm osc}}
   \right)
   \delta_{\rm eff} 
   \,,
\end{eqnarray}
when the reheating process of inflation completes.  This lepton
asymmetry is partially converted~\cite{LGorg} into the baryon asymmetry
through the ``sphaleron'' effects~\cite{sphaleron}, since it is produced
before the electroweak phase transition at $T \simeq 10^2 \GeV$. The
present baryon asymmetry is given by~\cite{L-to-B}\footnote{In the
present analysis we neglect the relative sign between the produced
lepton and baryon asymmetries.}
\begin{eqnarray}
 \frac{n_B}{s} = \frac{8}{23}\frac{n_L}{s} 
  \,.
\end{eqnarray}
Thus, after all, the present baryon asymmetry is give by
\begin{eqnarray}
 \label{nBs-final}
  \frac{n_B}{s} = \frac{2}{69}
  \frac{M T_R}{M_*^2}
  \left(
   \frac{m_{3/2}}{H_{\rm osc}}
   \right)
   \delta_{\rm eff} 
   \,.
\end{eqnarray}
We see that the produced baryon asymmetry becomes larger as the scale
$M$ increases, i.e., as the neutrino mass $m_{\nu}$
decreases. Therefore, the relevant flat direction for the AD
leptogenesis corresponds to the first family field, i.e., $\phi/\sqrt{2}
= L_1 = H_u$.

\begin{figure}[t]
 \label{Fig-BA}
 \centerline{\psfig{figure=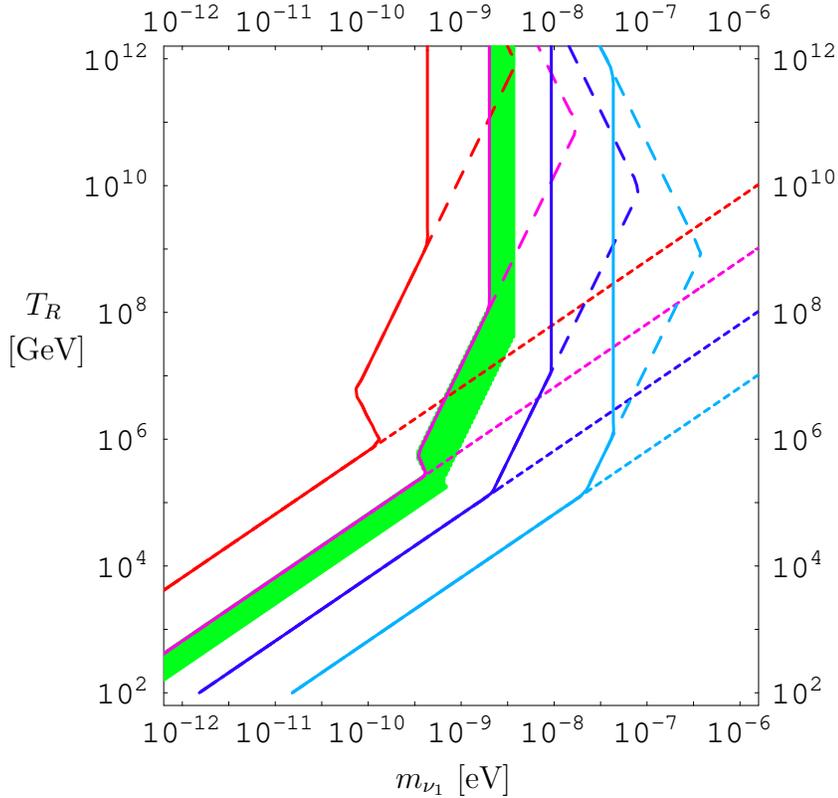,height=10cm}}
 \begin{picture}(0,0)
  \put(60,180){$T_R$}  
  \put(53,163){$[$GeV$]$}  
  \put(200,1){$m_{\nu_1}\,\,[$eV$]$}
 \end{picture}
 \caption{The contour plot of the baryon asymmetries $n_B/s$ in the
 $m_{\nu_1}$--$T_R$ plane.  The lines represent the contour plots for
 $n_B/s = 10^{-9}$, $10^{-10}$, $10^{-11}$, and $10^{-12}$ from the left
 to the right. The short-dashed lines represent the baryon asymmetry
 when one neglects the thermal effects. The long-dashed lines represent
 the ones including only the thermal mass $\propto T^2|\phi|^2$. The
 solid lines represent the baryon asymmetry including both thermal mass
 and $T^4\log(|\phi|^2)$ terms. The shaded region corresponds to the
 present baryon asymmetry, $n_B/s\simeq (0.4$--$1)\times 10^{-10}$. We
 have taken $\delta_{\rm eff}= 1$ in this figure. (See discussion in the
 text.)}
\end{figure}

Fig.~\ref{Fig-BA} shows the contour plot of the produced baryon
asymmetry in the $m_{\nu}$--$T_R$ plane. (Here we have used the relation
$m_{\nu} = \vev{H_u}^2 / M$.)
As shown in the figure, the present baryon asymmetry $n_B/s$ is
determined almost independently of the reheating temperature for a wide
range of $10^5 \lsim T_R\lsim 10^{12}$ GeV. In particular, for a
relatively high reheating temperature $10^8 \lsim T_R \lsim 10^{12}$ GeV,
the baryon asymmetry derived from the Eqs.~(\ref{HO1}) and
(\ref{nBs-final}) is given by the following simple from;
\begin{eqnarray}
 \frac{n_B}{s}
  \simeq
  10^{-11}
  \delta_{\rm eff}
  \times
  \left(
   \frac{m_{\nu_1}}{10^{-8}\,{\rm eV}}
   \right)^{-3/2}
   \left(
    \frac{m_{3/2} |a_m|}{1\,{\rm TeV}}
    \right)
   \,.
\end{eqnarray}
The reason why it is independent of the reheating temperature $T_R$ is
that the oscillation time $H_{\rm osc}$ is determined by the thermal
potential $T^4\log (|\phi|^2)$ in the higher temperature regime, and
$T_R$ dependence is canceled out in Eq.~(\ref{nBs-final}). Even in the
lower reheating temperature region $10^5 \lsim T_R \lsim 10^8$ GeV,
where $H_{\rm osc}$ is determined by the thermal-mass term potential
$T^2 |\phi|^2$, $T_R$ dependence is still mild, i.e., $n_B/s \propto
T_R^{1/3}$. The reheating-temperature independence discussed here is a
very attractive and remarkable feature of the present mechanism since
the produced baryon asymmetry crucially depends on $T_R$ in many other
baryogenesis scenarios.

In Fig.~\ref{Fig-BA}, we have taken $\delta_{\rm eff}= 1$. It is
expected that $\delta_{\rm eff}\sim {\cal O}(1)$, say, $\delta_{\rm
eff}\simeq 0.1$--$1$, unless there is an unnatural cancellation between
$\arg(a_m)$ and $\arg(a_H)$. Thus, the present baryon asymmetry in our
universe $n_B/s\simeq (0.4$--$1)\times 10^{-10}$ suggests an ultralight
neutrino of a mass $m_{\nu_1}\simeq (0.1$--$3)\times 10^{-9}$ eV for
$T_R\simeq 10^5$--$10^{12}$ GeV and $\delta_{\rm eff}\simeq
0.1$--$1$. We consider the region of $10^5 \lsim T_R \lsim 10^{12}$ GeV
throughout this paper, since it is the case for a large class of the
inflation models proposed in the framework of
supergravity~\cite{S-Inflations,S-chaotic}. In Sec.~\ref{sec-FN} we will
propose a model in which such an ultralight neutrino can be naturally
obtained.

%
\section{Prediction on the rate of neutrinoless double beta decay}
The neutrinoless double beta ($0\nu\beta\beta$) decay, if observed, is
the strongest evidence for the lepton number violation. The crucial
parameter to determine the $0\nu\beta\beta$ decay rate is $|m_{\nu_e
\nu_e}|\equiv |\sum_i U_{ei}^2 m_{\nu_i}|$, where $U_{\alpha i}$ is the
mixing matrix which diagonalize the neutrino mass matrix.\footnote{As
for general studies of the $|m_{\nu_e \nu_e}|$ using the neutrino masses
and mixings, see, for example, Ref.~\cite{MeeStudy} and references
therein.}  If the mass of the lightest neutrino is actually so small,
$m_{\nu_1}\sim 10^{-9}$ eV, as discussed in the previous sections, the
contribution from $m_{\nu_1}$ to $m_{\nu_e \nu_e}$ can be completely
neglected and hence the parameter $|m_{\nu_e \nu_e}|$ is written in
terms of masses and mixings of two other neutrinos as
\begin{eqnarray}
 |m_{\nu_e \nu_e}| = |U_{e2}^2 m_{\nu_2} + U_{e3}^2 m_{\nu_3}| 
  \,.
\end{eqnarray}
Therefore, the $|m_{\nu_e \nu_e}|$ is determined by $U_{e3}$ and the
parameters of atmospheric and solar neutrino oscillations; $\delta
m_{\rm atm}^2 \simeq m_{\nu_3}^2$, $\delta m_{\rm sol}^2 \simeq
m_{\nu_2}^2$ and $\tan^2\theta_{\rm sol} \equiv
|U_{e2}/U_{e1}|^2$. Namely, it is given by
\begin{eqnarray}
 |m_{\nu_e \nu_e}| \simeq
  \left|
   \left( 1 - |U_{e3}|^2 \right)
   \sin^2 \theta_{\rm sol}
   \sqrt{\delta m_{\rm sol}^2}
   +
   |U_{e3}|^2 
   e^{i\alpha}
   \sqrt{\delta m_{\rm atm}^2}
   \,\,
   \right|
   \,,
\end{eqnarray}
where $\alpha$ denotes the relative phase between the two terms.

\begin{figure}[t]
 \centerline{
 {\psfig{figure=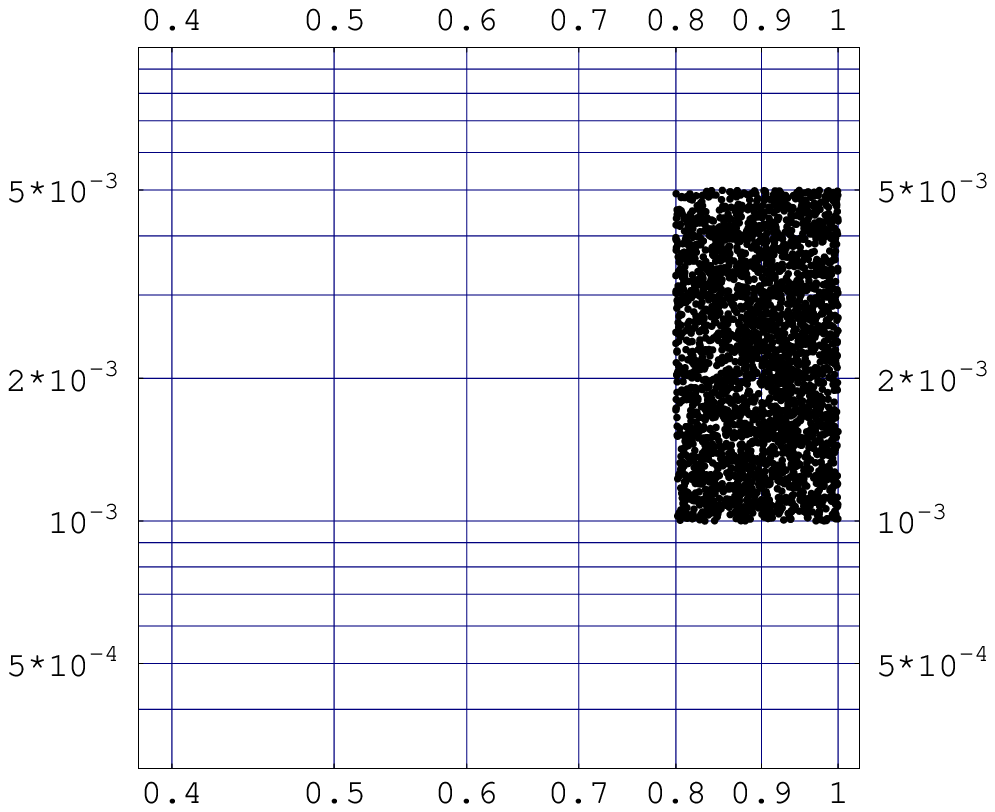,height=7cm}} 
 }
 \begin{picture}(0,0)
  \put(70,140){$\delta m_{\rm atm}^2$}  
  \put(70,123){$[$eV$^2]$}  
  \put(200,1){$\sin^2 2\theta_{\rm atm}$}
 \end{picture}
 \caption{The parameter range we have taken for the atmospheric neutrino
 oscillation.}
 \label{fig-atm-range}
 \vspace{1cm}
 \centerline{ 
 {\psfig{figure=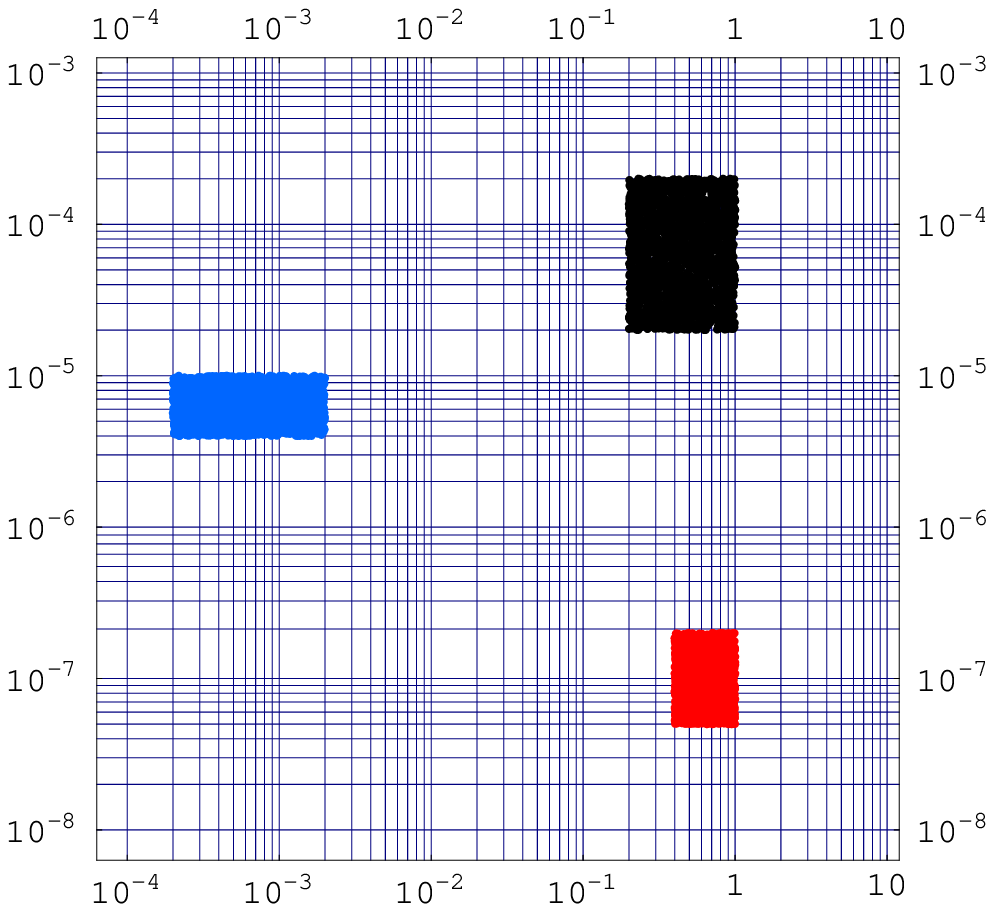,height=7cm}}
 }
 \begin{picture}(0,0)
  \put(80,140){$\delta m_{\rm sol}^2$}  
  \put(80,123){$[$eV$^2]$}  
  \put(200,1){$\tan^2 \theta_{\rm sol}$}
 \end{picture}
 \caption{The parameter ranges we have taken for the large angle MSW,
 the small angle MSW and the LOW solutions to the solar neutrino
 problem.}
 \label{fig-sol-range}
\end{figure}

\begin{figure}[t]
 \centerline{ {\psfig{figure=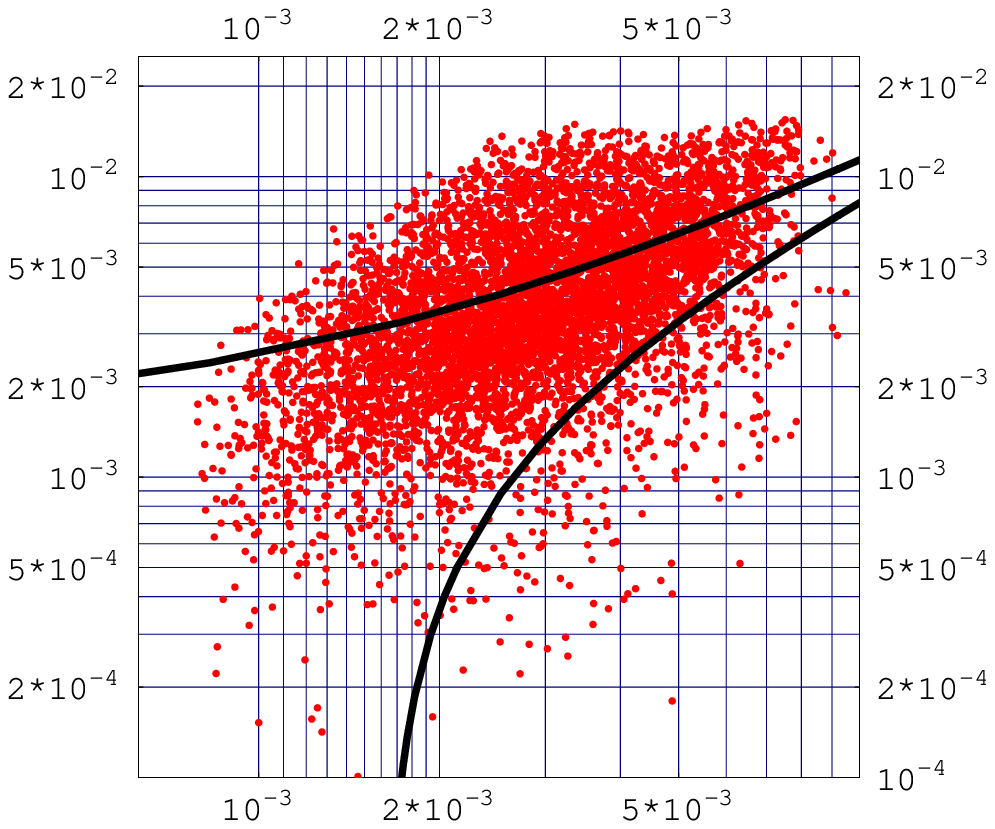,height=8cm}} } 
 \begin{picture}(0,0)
  \put(45,140){$|m_{\nu_e \nu_e}|$}  
  \put(50,123){$[$eV$]$}  
  \put(180,1)
  {$\sin^2\theta_{\rm sol} \sqrt{\delta m_{\rm sol}^2}\,\,\,[$eV$]$}
 \end{picture}
 \caption{The predicted value of $|m_{\nu_e \nu_e}|$ for the large angle
 MSW solution. The solid lines represent the upper and lower values of
 $|m_{\nu_e \nu_e}|$ for $m_{\nu_1}\simeq 0$. The plots represent the
 values for the case when the $m_{\nu_1}$ is allowed to be $m_{\nu_1}
 \le (1/\sqrt{2}) m_{\nu_2}$. Here, we have required $|U_{e3}| < 0.15$,
 to satisfy the CHOOZ limit.}
 \label{fig-mee-LMA-015}
 \vspace{1cm}
 \centerline{ {\psfig{figure=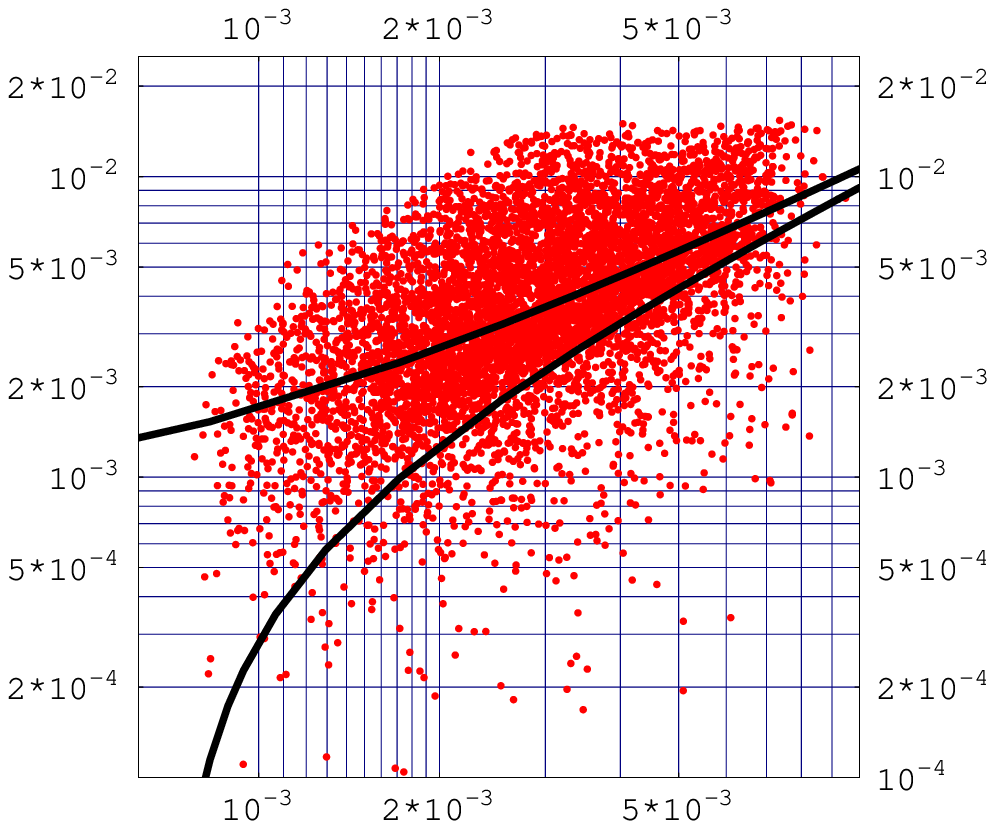,height=8cm}} } 
 \begin{picture}(0,0)
  \put(45,140){$|m_{\nu_e \nu_e}|$}  
  \put(50,123){$[$eV$]$}  
  \put(180,1)
  {$\sin^2\theta_{\rm sol} \sqrt{\delta m_{\rm sol}^2}\,\,\,[$eV$]$}
 \end{picture}
 \caption{Same as Fig.~\ref{fig-mee-LMA-015}, but for $|U_{e3}| < 0.10$.}
 \label{fig-mee-LMA-010}
\end{figure}

We calculate the predicted value of $|m_{\nu_e \nu_e}|$ for the large
angle MSW, the small angle MSW and the LOW solutions, taking the
parameters allowed for atmospheric and solar neutrino oscillations that
are shown in Fig.~\ref{fig-atm-range}~\cite{SK-Atm} and
Fig~\ref{fig-sol-range}~\cite{Lisi}.  In the case of large angle MSW
solution, $|m_{\nu_e \nu_e}|$ is sensitive mainly to the parameter of
the solar neutrino oscillation, $\sin^2 \theta_{\rm sol} \sqrt{\delta
m_{\rm sol}^2}$. The predicted value of $|m_{\nu_e \nu_e}|$ for the
large angle MSW solution is shown in Fig.~\ref{fig-mee-LMA-015}.  Here,
we have required $|U_{e3}| < 0.15$ from the CHOOZ
experiment~\cite{CHOOZ}.  For a comparison, we also show the possible
values of $|m_{\nu_e \nu_e}|$ when we allow $m_{\nu_1}$ to be relatively
large as $m_{\nu_1} \le (1/\sqrt{2}) m_{\nu_2}$. We see from
Fig.~\ref{fig-mee-LMA-015} that the $|m_{\nu_e \nu_e}|$ is predicted in
a narrow range.  It is very encouraging that the predicted $|m_{\nu_e
\nu_e}|$ can be large enough to be accessible at future experiments such
as GENIUS~\cite{GENIUS}.  Furthermore, if the $U_{e3}$ becomes more
constrained by future experiments~\cite{ue3exp}, $|m_{\nu_e \nu_e}|$ is
predicted in a much narrower range as shown in
Fig.~\ref{fig-mee-LMA-010}, where we have required $|U_{e3}| < 0.10$.

\begin{figure}[t]
 \centerline{ {\psfig{figure=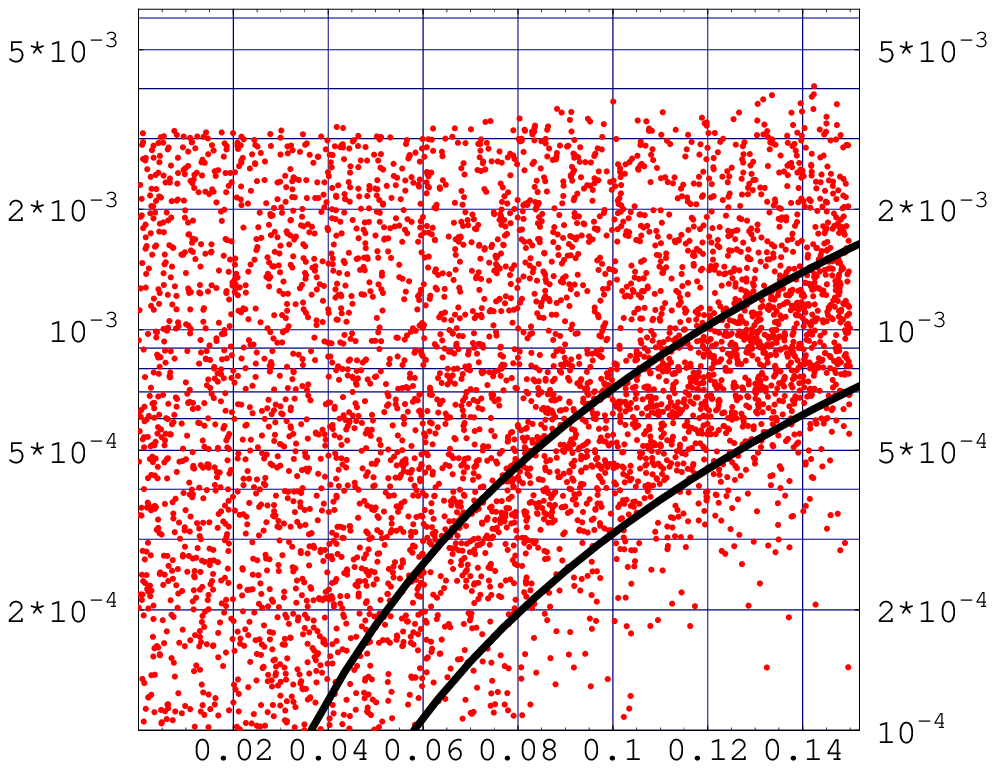,height=8cm}} } 
 \begin{picture}(0,0)
  \put(45,140){$|m_{\nu_e \nu_e}|$}  
  \put(50,123){$[$eV$]$}  
  \put(210,1){$|U_{e3}|$}
 \end{picture}
 \caption{The predicted value of $|m_{\nu_e \nu_e}|$ for the small angle
 MSW solution. The solid lines represent the upper and lower values of
 $|m_{\nu_e \nu_e}|$ for $m_{\nu_1}\simeq 0$. The plots represent the
 values for the case when the $m_{\nu_1}$ is allowed to be $m_{\nu_1}
 \le (1/\sqrt{2}) m_{\nu_2}$.}
 \label{fig-mee-SMA}
 \vspace{1cm}
 \centerline{ {\psfig{figure=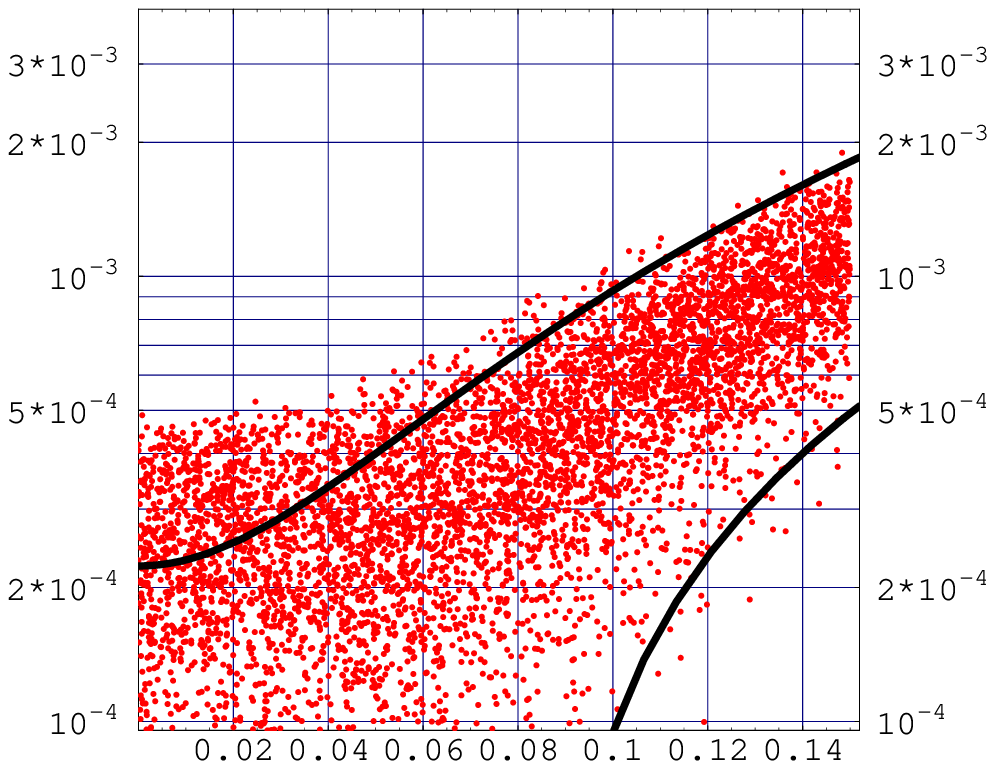,height=8cm}} } 
 \begin{picture}(0,0)
  \put(45,140){$|m_{\nu_e \nu_e}|$}  
  \put(50,123){$[$eV$]$}  
  \put(210,1){$|U_{e3}|$}
 \end{picture}
 \caption{Same as Fig.~\ref{fig-mee-SMA}, but for the LOW solution.}
 \label{fig-mee-LOW}
\end{figure}

On the other hand, the $|m_{\nu_e \nu_e}|$ is sensitive to $|U_{e3}|$ in
the case of the small angle MSW and the LOW solutions. The results are
shown in Fig.~\ref{fig-mee-SMA} and \ref{fig-mee-LOW}. Because
$|U_{e3}|$ is highly constrained by the CHOOZ experiment, the predicted
value of $|m_{\nu_e \nu_e}|$ is so small. Even in these cases the
contribution from $m_{\nu_1}$ can not enhance $|m_{\nu_e \nu_e}|$
because it is too small.

In all cases, the $|m_{\nu_e \nu_e}|$ is predicted with high accuracy,
depending on the solar and atmospheric neutrino oscillation parameters
and $U_{e3}$. Therefore, the presence of such an ultralight neutrino
indicated from the present baryon asymmetry can be tested at near future
experiments. However, notice that the results shown in this section is a
generic consequence of the mass hierarchy $m_{\nu_1} \ll m_{\nu_{2,3}}$.
Thus, we consider that the $0\nu\beta\beta$ decay provides only a
consistency test for our hypothesis $m_{\nu_1}\sim 10^{-9}$ eV.


\section{A model for the ultralight neutrino}
\label{sec-FN}
%
In Sec.~\ref{sec-AD} we have shown that the baryon asymmetry in the
present universe predicts the mass for the lightest neutrino in a narrow
region, $m_{\nu_1}\simeq (0.1$--$3)\times 10^{-9}$ eV. Together with
neutrino masses required to solve the solar and atmospheric neutrino
anomalies, this suggests a very large mass hierarchy between the
lightest and the heavier two neutrinos. In this section we show an
explicit model based on a Froggatt-Nielsen (FN) mechanism~\cite{FroNie},
in which such an large mass hierarchy is naturally obtained.

We adopt a discrete $Z_6$ as the FN symmetry instead of a continuous
U$(1)_{\rm FN}$. We see that the discrete symmetry is crucial to produce
the required large mass hierarchy in the neutrino sector. Before
describing our model, we briefly review the FN model~\cite{FroNie},
which explains the observed hierarchies in quark and lepton mass
matrices. This model is based on a U$(1)_{\rm FN}$ symmetry that is 
broken by the vacuum-expectation value of $\Phi$, $\vev{\Phi} \neq 0$. 
Here $\Phi$ is a gauge singlet FN field carrying the FN charge 
$Q_{\Phi} = -1$. Then, all Yukawa couplings arise from nonrenormalizable 
interactions of $\Phi$ and are given by the following form;
\begin{eqnarray}
 W &=& h_{ij} 
  \left(
   \frac{\vev{\Phi}}{M_*}
   \right)^{Q_i + Q_j}
   \Psi_i \Psi_j H_{u(d)}
   \nonumber \\
 &=&
  h_{ij}
  \,\,\ep^{Q_i + Q_j}\,\,
  \Psi_i \Psi_j H_{u(d)} \,,
\end{eqnarray}
where $h_{ij}$ are ${\cal O}(1)$ coupling constants, $Q_i$ are the FN
charges of various chiral superfields $\Psi_i$ and $\ep \equiv
\vev{\Phi}/M_*$. The observed mass hierarchies for quarks and charged
leptons are well explained by taking suitable FN charges for them. For
instance, we assign FN charges ($a$, $a$, $a+1$) for lepton doublets
$L_i$, while giving charges ($0$, $1$, $2$) to the right-handed charged
leptons $E_i$, with $\ep \simeq
0.05$--$0.1$~\cite{FN001,SatoYana-paper}. Here, we take $a = 0$ or
$1$~\cite{FN001}. Charges for the quarks are determined if one assumes
the SU$(5)$ grand unified theory~\cite{FN001}.

The mass matrix for the neutrinos in this model is determined by the FN
charges for the lepton doublets $L_i$~\cite{FN001}. To see this we first
discuss the mass matrix for the heavy right-handed neutrinos $N_i$,
which is given by;
\begin{eqnarray}
 M_{R\,ij} = g_{ij}\,\,\ep^{Q_i + Q_j}\,\,M_0 \,,
\end{eqnarray}
where $M_0$ represents some right-handed neutrino mass scale and
$g_{ij}$ are coupling constants of ${\cal O}(1)$ like $h_{ij}$.
Hereafter, we will take a basis where the mass matrix for the charged
leptons is diagonal.\footnote{One might wonder if the mixing matrix from
the charged lepton sector would change the discussion above, since the
mass matrix for the charged leptons has off-diagonal elements in the FN
mechanism. However, the correction from this effect yields higher order
terms in $\ep$, and hence we can safely neglect it.} The charges for the
lepton doublets $L_i$ and right-handed neutrinos $N_i$ are listed in
Table.~\ref{tableFN}.
\begin{table}[t]
 \begin{center}
   \begin{tabular}{|c|c c c c c c|}
    \hline
    $\Psi_i$ & $L_3$ & $L_2$ & $L_1$ & $N_3$ & $N_2$ & $N_1$ 
    \\ \hline
    $Q_i$ & $a$ & $a$ & $a+1$ & $b$ & $c$ & $d$
    \\ \hline
   \end{tabular}
 \end{center}
 \caption{The FN charges of lepton doublets and right-handed
 neutrinos. $a = 0$ or $1$. We assume, for simplicity, $0\le b\le c <
 d$.}
 \label{tableFN}
\end{table}
Then, the neutrino Dirac mass matrix $m_D$ and the
right-handed neutrino mass matrix $M_R$ are given by the following forms;
\begin{eqnarray}
 m_D &=&
  \vev{H_u}
  \left(
   \begin{array}{ccc}
    \ep^{a+1} & 0 & 0
     \\
    0 & \ep^a & 0
     \\
    0 & 0 & \ep^a
   \end{array}
   \right)
   \left(
    \begin{array}{ccc}
     h_{11} & h_{12} & h_{13}
      \\
     h_{21} & h_{22} & h_{23}
      \\
     h_{31} & h_{32} & h_{33}
    \end{array}
    \right)
    \left(
     \begin{array}{ccc}
      \ep^d & 0 & 0
       \\
      0 & \ep^c & 0
       \\
      0 & 0 & \ep^b
     \end{array}
     \right)\,,
     \nonumber \\
 M_R &=&
  M_0
  \left(
   \begin{array}{ccc}
    \ep^d & 0 & 0
     \\
    0 & \ep^c & 0
     \\
    0 & 0 & \ep^b
   \end{array}
   \right)
   \left(
    \begin{array}{ccc}
     g_{11} & g_{12} & g_{13}
      \\
     g_{12} & g_{22} & g_{23}
      \\
     g_{13} & g_{23} & g_{33}
    \end{array}
    \right)
    \left(
     \begin{array}{ccc}
      \ep^d & 0 & 0
       \\
      0 & \ep^c & 0
       \\
      0 & 0 & \ep^b
     \end{array}
     \right)\,.
\end{eqnarray}
We obtain the neutrino mass matrix as
\begin{eqnarray}
 \label{mnu-matrix}
  m_{\nu} &=&
  m_D \frac{1}{M_R} m_D^T
  \nonumber \\
 &=&
  \frac{\ep^{2a} \vev{H_u}}{M_0}
  \left(
   \begin{array}{ccc}
    \ep & 0 & 0
     \\
    0 & 1 & 0
     \\
    0 & 0 & 1
   \end{array}
   \right)
   \left(
    \,\{h_{ij}\}\,
    \right)
    \left(
     \,\{g_{ij}\}\,
     \right)^{-1}
     \left(
      \,\{h_{ij}\}\,
      \right)^T
      \left(
       \begin{array}{ccc}
	\ep & 0 & 0
	 \\
	0 & 1 & 0
	 \\
	0 & 0 & 1
       \end{array}
       \right)
       \nonumber \\
 &\sim&
  \frac{\ep^{2a} \vev{H_u}}{M_0}
  \left(
   \begin{array}{ccc}
    \ep^2 & \ep & \ep
     \\
    \ep & 1 & 1
     \\
    \ep & 1 & 1
   \end{array}
   \right)\,.
\end{eqnarray}
As shown in Ref.~\cite{FN001}, this mass matrix can naturally lead to a
large $\nu_{\mu}$--$\nu_{\tau}$ mixing angle, which is suggested from
the atmospheric neutrino oscillation~\cite{SK-Atm}. It is very crucial
that the FN charges of the right-handed neutrinos are completely
canceled out in the neutrino mass matrix in Eq.~(\ref{mnu-matrix}) and
hence the masses of the neutrinos are determined only by the charges of
the lepton doublets, ($a$, $a$, $a+1$). This gives a mild mass
hierarchy,
\begin{eqnarray} 
m_{\nu _3}:m_{\nu _2}:m_{\nu _1} \simeq 1:1:\epsilon ^2 \simeq {\cal O}(1):
{\cal O}(1): {\cal O}(10^{-2}).
\end{eqnarray}

Now let us turn to our FN model. To change the above point, we suppose 
that the broken FN symmetry is not a U$(1)_{\rm FN}$ but a discrete 
symmetry $Z_n$ with $n = 2 d$. Then, the mass matrix for the
right-handed neutrino $M_R$ changes into the following form;
\begin{eqnarray}
 M_R =
  M_0
  \left(
   \begin{array}{ccc}
    g_{11} & g_{12}\,\ep^{c+d} & g_{13}\,\ep^{b+d}
     \\
    g_{12}\,\ep^{c+d} & g_{22}\,\ep^{2 c} & g_{23}\,\ep^{b+c}
     \\
    g_{13}\,\ep^{b+d} & g_{23}\,\ep^{b+c} & g_{33}\,\ep^{2 b}
   \end{array}
   \right)\,.
   \label{matMR}
\end{eqnarray}
Here we assume $0 \le b \le c < d$. Notice that the Majorana mass for
$N_1$ is no longer suppressed by the power of $\epsilon$, which is a
basic point to yield an extremely small neutrino mass $m_{\nu _
1}$. However, the structure of the neutrino mass matrix looks similar to
the original one;
\begin{eqnarray}
 m_{\nu}  &=&
  \frac{\ep^{2a} \vev{H_u}}{M_0}
  \left(
   \begin{array}{ccc}
    \ep & 0 & 0
     \\
    0 & 1 & 0
     \\
    0 & 0 & 1
   \end{array}
   \right)
   \left(
    \,\{h_{ij}\}\,
    \right)
    \left(
     \begin{array}{ccc}
      g_{11}\,\ep^{-2d} & g_{12} & g_{13}
       \\
      g_{12} & g_{22} & g_{23}
       \\
      g_{13} & g_{23} & g_{33}
     \end{array}
     \right)^{-1}
     \left(
      \,\{h_{ij}\}\,
      \right)^T
      \left(
       \begin{array}{ccc}
	\ep & 0 & 0
	 \\
	0 & 1 & 0
	 \\
	0 & 0 & 1
       \end{array}
       \right)
       \nonumber \\
 &\sim&
  \frac{\ep^{2a} \vev{H_u}}{M_0}
  \left(
   \begin{array}{ccc}
    \ep^2 & \ep & \ep
     \\
    \ep & 1 & 1
     \\
    \ep & 1 & 1
   \end{array}
   \right)\,.
\end{eqnarray}
We see that one of the mass eigenvalue of this mass matrix strongly
suppressed as $\sim \ep^{2(1+d)}$. This suppression is also understood
directly by taking the determinant of the above mass matrix. For the
mass hierarchy required from the successful AD leptogenesis, it is
suitable to take $d=3$ ($Z_6$).

To demonstrate our point, we randomly generate ${\cal O}(1)$ couplings
$h_{ij}$ and $g_{ij}$. Namely we calculate the mass matrix for
neutrinos, taking the magnitudes of the couplings $h_{ij}$ and $g_{ij}$
to be in a range $0.5$ -- $1.5$ and their phases to be $0$ -- $2\pi$.
We also take $\ep = 0.05$--$0.1$ randomly.\footnote{A similar
calculation was done in Ref.~\cite{SatoYana-paper}, where they adopted
the U$(1)_{\rm FN}$ model.}  Here, we have required the parameters
$r\equiv \delta m^2_{\rm sol} /\delta m^2_{\rm atm} = (m_{\nu_3}^2-
m_{\nu_2}^2)/(m_{\nu_2}^2- m_{\nu_1}^2)$, $\sin^2 2\theta_{\rm
atm}\equiv 4|U_{\mu 3}|^2\left(1 - |U_{\mu 3}|^2 \right)$ and
$\tan^2\theta_{\rm sol}\equiv |U_{e2}/U_{e1}|^2$ to be consistent with
the parameter regions shown in Figs.~\ref{fig-atm-range} and
\ref{fig-sol-range}.  We have here required $|U_{e3}| < 0.15$ to satisfy
CHOOZ limit~\cite{CHOOZ}.
\begin{figure}[t]
 \centerline{ {\psfig{figure=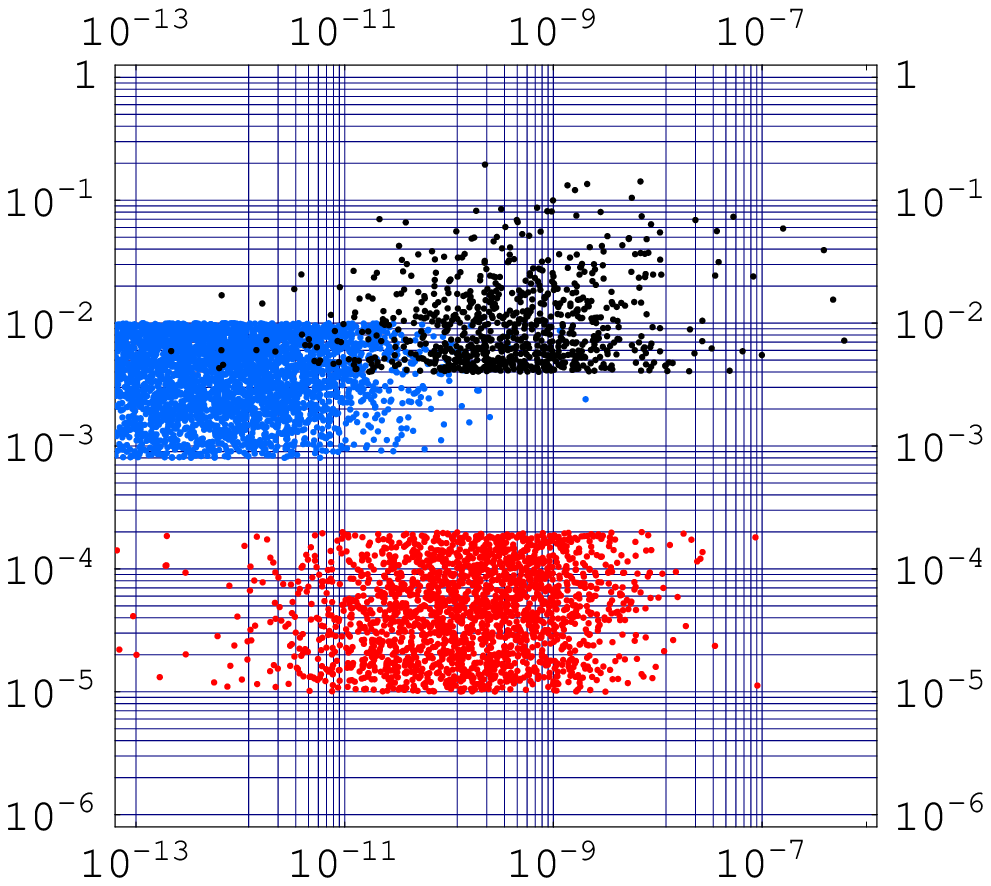,height=9.5cm}} }
 \begin{picture}(0,0)
  \put(17,140){$r = \displaystyle\frac{\delta m^2_{\rm sol}}
  {\delta m^2_{\rm atm}}$}
  \put(210,1){$m_{\nu_1}\,\,[$eV$]$}
 \end{picture}
 \caption{The plot for $r = \delta m^2_{\rm sol} /\delta m^2_{\rm atm}$
 and $m_{\nu_1}$ in a Froggatt-Nielsen model with a discrete $Z_6$
 symmetry.}
 \label{Fig-mass}
\end{figure}
Fig.~\ref{Fig-mass} shows the obtained mass of the lightest neutrino,
$m_{\nu_1}$. We can see that, an ultralight neutrino of mass
$m_{\nu_1}\simeq (0.1$--$3) \times 10^{-9}$ eV is naturally obtained.

%
%
\section{Discussion and conclusions}
In this paper we have performed a reanalysis on Affleck-Dine (AD)
leptogenesis taking into account of all the relevant thermal effects.
Then, we have pointed out that the baryon asymmetry is produced almost
independently of the reheating temperature $T_R$ and it is determined
mainly by the mass of the lightest neutrino $m_{\nu_1}$ in a wide range
of the reheating temperature $T_R\sim 10^5$--$10^{12}$ GeV.  Notice that
such reheating temperatures, $T_R\sim 10^5$--$10^{12}$ GeV, are
naturally realized in a large class of SUSY inflation
models~\cite{S-Inflations,S-chaotic}. This reheating-temperature
independence is a very attractive feature of the AD leptogenesis.

Furthermore, we have shown that the present baryon asymmetry predicts
the mass for the lightest neutrino in a very narrow region,
$m_{\nu_1}\simeq (0.3$--$1)\times 10^{-9}$ eV.  We have also proposed an
explicit model based on a Froggatt-Nielsen mechanism~\cite{FroNie} with
a discrete symmetry $Z_6$, where such an ultralight neutrino indicated
from the AD leptogenesis is naturally obtained.

Such a small mass of the lightest neutrino means that the mass parameter
$m_{\nu_e \nu_e}$ contributing to the $0\nu\beta\beta$ decay is
determined by the masses and mixings of two other neutrinos. Actually,
we have shown that $|m_{\nu_e \nu_e}|$ can be predicted with high
accuracy, by observable neutrino oscillation parameters such as $\delta
m_{\rm atm}^2$, $\delta m_{\rm sol}^2$, $\sin^2 \theta_{\rm sol}$ and
$U_{e3}$. In particular, when the large angle MSW solution is the case
and $\sin^2 \theta_{\rm sol} \sqrt{\delta m_{\rm sol}^2}$ is relatively
large, the value of $|m_{\nu_e \nu_e}|$ is predicted as $|m_{\nu_e
\nu_e}|\simeq 10^{-3}$--$10^{-2}$ eV, which may be testable at future
$0\nu\beta\beta$ decay experiments such as GENIUS~\cite{GENIUS}.

\section*{Acknowledgements}
The work of K.H. was supported by the Japanese
Society for the Promotion of Science. T.Y. acknowledges partial
support from the Grant-in-Aid for Scientific Research from the
Ministry of Education, Sports, and Culture of Japan, on Priority Aerea
\# 707: ``Supersymmetry and Unified Theory of Elementary Particles''.
%


%

%
%

\end{document}